\input harvmac
\sequentialequations
%%%%%%%%%%%%%%%%%%%%%%%%%%%definitions%%%%%%%%%%%%%%%%%%%%%%%%%%%%%%%
\def\coshi{\cosh\delta_i}
\def\sinhi{\sinh\delta_i}
\def\reff{R_{eff}}
\def\vin{V_{in}}
\def\sabs{\sigma_{abs}}
\def\sbabs{\bar{\sigma}_{abs}}
\def\ie{{\it i.e.}}
\def\phinf{\phi_\infty}
%%%%%%%%%%%%%%%%%%%%%%%%%%%references%%%%%%%%%%%%%%%%%%%%%%%%%%%%%%%

%%%%%%%%%%%%%%%%%%%%%%%%%%%%references%%%%%%%%%%%%%%%%%%%%
\lref\horowitz{G.T. Horowitz, {\it The Origin of Black Hole Entropy in 
String Theory}, to appear in the proceedings of the Pacific Conference 
on Gravity
and Cosmology, gr-qc/9604051 .}
\lref\maldacena{J. Maldcena, {\it Black Holes in String Theory}, 
hep-th/9607235 .}
\lref\polchinski{J. Polchinski, hep-th/9510017 .}
\lref\strominger{A. Strominger and C. Vafa, hep-th/9601029 .}
\lref\cy{M. Cvetic and D. Youm, 
{\it General Rotating, Five Dimensional Black Holes of Toroidally 
Compactified Heterotic String}, 
Nucl. Phys. {\bf B476} (1996) 118, hep-th/9603100 .} 
\lref\dmthree{S.R. Das and S.D. Mathur, {\it Interactions Involving D-Branes},
hep-th/???}
\lref\kol{B. Kol and A. Rajaraman, {\it Fixed Scalars and Suppresion of Hawking
Radiation}, Phys. Rev. {\bf D56} (1997) 983, hep-th/9608126 .}
\lref\igor{S.S. Gubser and I.R. Klebanov, {\it Emission of Charged 
Particles from Four-
and Five-Dimensional Black Holes}, Nucl. Phys. {\bf B482} (1996) 173, 
hep-th/9608108 .}
\lref\dkt{H.F. Dowker, D. Kastor and J.Traschen, {\it U-Duality, D-Branes and
Black Hole Emission Rates: Agreements and Disagreements}, hep-th/9702109.}
\lref\callan{C.G. Callan and J. Maldacena, {\it D-Brane Approach
to Black Hole Quantum Mechanics}, Nucl. Phys. {\bf B472} (1996) 591, 
hep-th/9602043 .} 
\lref\hs{G. Horowitz and A. Strominger, {\it Counting States of
Near Extremal Black Holes}, hep-th/9602051 .}  
\lref\hms{G.T. Horowitz, J. Maldacena and A. Strominger, 
{\it Nonextremal Black Hole Microstates and U-Duality}, Phys. Lett. 
{\bf 383B} (1996) 151, 
hep-th/9603109.}
\lref\dmone{S.R. Das and S.D. Mathur, {\it Excitations of D-Strings, Entropy
and Duality}, Phys. Lett. {\bf 375B} (1996) 103, hep-th/9601152 .}
\lref\wadia{A. Dhar, G. Mandal and S.R. Wadia,  {\it Absorption vs. Decay of
Black Holes in String Theory and T-Symmetry}, Phys. Lett. {\bf 388B} (1996) 51, 
hep-th/9605234 .}
\lref\msuss{J. Maldacena and L. Susskind, {\it D-Branes and Fat Black
Holes}, Nucl. Phys. {\bf B475} (1996) 670, hep-th/9604042 .}  
\lref\dmtwo{S.R. Das and S.D. Mathur. {\it Comparing
Decay Rates for Black Holes and D-Branes}, Nucl. Phys. {\bf B478} (1996) 561, 
hep-th/9606185 .}
\lref\gibbons{G.W. Gibbons, {\it Vacuum Polarization and the Spontaneous Loss
of Charge by Black Holes}, Comm. Math. Phys. {\bf 44}, 245 (1975).}
\lref\hashimoto{A. Hashimoto and I.R. Klebanov, {\it Decay of Excited D-branes}, 
Phys. Lett. {\bf 381B} (1996) 437, hep-th/9604065 .}
\lref\gubser{S.S. Gubser and I.R. Klebanov, {\it Four-Dimensional Greybody
Factors and the Effective String}, Phys. Rev. Lett. {\bf 77}, 4491, hep-th/9609076 .}
\lref\fixed{C.G. Callan, S.S. Gubser, I.R. Klebanov and A.A. Tseytlin, 
{\it Absorption of Fixed Scalars and the D-brane Approach to Black Holes},
Nucl. Phys. {\bf B489} (1997) 65, hep-th/9610172 .}
\lref\gub{S.S. Gubser, {\it Can the Effective String See Higher Partial Waves},
hep-th/9704195 .}
\lref\dealwis{S.P. de Alwis and K. Sato, {\it Radiation from a Class of String
Theoretic Black Holes}, hep-th/9611189 .}
\lref\fixedtwo{I.R. Klebanov and M. Krasnitz, {\it Fixed Scalar Greybody
Factors in Five and Four Dimensions}, Phys. Rev. {\bf D55} (1997) 3250, 
hep-th/9612051 .}
\lref\esko{E. Keski-Vakkuri and P. Krauss, {\it Microcanonical D-branes and
Back Reaction}, hep-th/9610045 .}
\lref\hashimotob{A. Hashimoto, {\it Perturbative Dynamics of Fractional Strings
on Multiply Wound D-Strings}, hep-th/9610250 .}
\lref\polhemus{G. Polhemus, {\it Statistical Mechanics of Multiply Wound
D-Branes}, hep-th/9612130 .}
\lref\kk{M. Krasnitz and I.R. Klebanov, {\it Testing Effective String Models of Black
Holes with Fixed Scalars}, hep-th/9703216 .}
\lref\page{D.N. Page, {\it Particle Emission Rates from a Black Hole: 
Massless Particles from an 
Uncharged Hole}, Phys. Rev. {\bf D13} (1976) 198.}
\lref\unruh{W.G. Unruh, {\it Absorption Cross Section of Small Black Holes}, 
Phys. Rev. {\bf D14} 
(1976) 325.}
\lref\star{A.A. Starobinski and S.M. Churilov, {\it Amplification of 
Electromagnetic and Gravitational 
Waves by A Rotating ``Black Hole''}, Sov. Phys. JETP {\bf 38} (1974) 1.}
\lref\press{S.A. Teukolsky and W.H. Press, {\it Perturbations of a Rotating Black 
Hole III: Interaction of the Hole with Gravitational and Electromagnetic Radiation}, 
Ap. J. {\bf 193} (1974) 443.}
\lref\hawk{S.W. Hawking, {\it Particle Creation by Black Holes}, 
Comm. Math. Phys. {\bf 43} (1975) 199.}
\lref\ms{J. Maldacena and A. Strominger, {\it Black Hole Greybody Factors and
D-Brane Spectroscopy}, Phys. Rev. {\bf D55} (1997) 861, hep-th/9609026 .} 
\lref\sa{M. Abramowitz and I.A. Stegun, {\it Handbook of Mathematical
Functions}, National Bureau of Standards New York: Wiley (1964).}
\lref\fluids{M. Van Dyke, {\it Perturbation Methods in Fluid Mechanics},
The Parabolic Press, Stanford California (1975).}
\lref\km{I.R. Klebanov and S.D. Mathur, {\it Black Hole Greybody Factors and Absorption of 
Scalars by Effective Strings}, hep-th/9701187.}
\lref\mstwo{J. Maldacena and A. Strominger, {\it Universal Low-Energy Dynamics for Rotating 
Black Holes}, hep-th/9702015.}
\lref\hawking{S.W. Hawking and M.M. Taylor-Robinson, {\it Evolution of 
Near Extremal Black Holes}, Phys.Rev. {\bf D55} (1997) 7680, 
hep-th/9702045.}
\lref\dasgupta{S. Das, A. Dasgupta and T. Sarkar, {\it High Energy Effects on D-Brane and 
Black Hole Emission Rates}, hep-th/9702075}
\lref\dgm{S.R. Das, G. Gibbons and S.D. Mathur, {\it Universality of Low-Energy 
Absorption Cross-Sections
for Black Holes}, Phys. Rev. Lett. {\bf 78} (1997) 417, hep-th/9609052 .}
\lref\larsen{F. Larsen, {\it A String Model of Black Hole Microstates}, Phys. Rev {\bf D56}
(1997) 1005, hep-th/9702153 .} 
\lref\clone{M. Cvetic and F. Larsen, {\it General Rotating
Black Holes in String Theory: Grey Body Factors and Event Horizons}, hep-th/9705192 .}
\lref\cltwo{M. Cvetic and F. Larsen, {\it Grey Body Factors For Rotating Black Holes in 
Four Dimensions}, hep-th/9706071 .}
\lref\marika{M. Taylor-Robinson, {\it Absorption of Fixed Scalars}, hep-th/9704172 .}
\lref\krt{I.R. Klebanov, A. Rajaraman and A.A. Tseytlin, {\it Intermediate Scalars and
the Effective String Model of Black Holes}, hep-th/9704112 .}
\lref\gubthree{S.S. Gubser, {\it Absorption of Photons and Fermions by Black Holes in 
Four-Dimensions}, hep-th/9706100 .}

%%%%%%%%%%%%%%%%%%%%%%%%%%%References%%%%%%%%%%%%%%%%%%%%%%%%%%%%%%%%%%
%%%%%%%%%%%%%%%%%%%%%%%%%%% Title Page %%%%%%%%%%%%%%%%%%%%%%%%%%%%%%%

\Title{\vbox{\baselineskip12pt
\hbox{UMHEP-444}
\hbox{hep-th/9707157}}}
{\vbox{\centerline{\titlerm A Very Effective String Model?}
 }}
{\baselineskip=12pt
\centerline{David Kastor$^{a}$
and Jennie Traschen$^{b}$  }
\bigskip
\centerline{\sl Department of Physics and Astronomy}
\centerline{\sl University of Massachusetts}
\centerline{\sl Amherst, MA 01003-4525}
\centerline{\it ${}^{a}$ Internet: kastor@phast.umass.edu}
\centerline{\it ${}^{b}$ Internet: lboo@phast.umass.edu}}
\bigskip
\medskip
\centerline{\bf Abstract}
\bigskip
Additional evidence is presented for a recently proposed effective string model,
conjectured to hold throughout the parameter space of the basic 5
dimensional, triply charged black  holes, which includes the effects of 
brane excitations, as well as momentum modes. 
We compute the low energy spacetime absorption coefficient $\sigma$ for the
scattering of a triply-charged scalar field in the near extremal case, and
conjecture an exact form  for $\sigma$. It is shown that this form of $\sigma$ arises
simply from  the effective string model. This agreement encompasses both
statistical factors coming  from the Bose distributions of string excitations
and a prefactor which depends on the  effective string radius. An interesting
feature of the effective string model is that the change in mass of the
effective string system in an emission process is not equal to  the change in
the energies of the effective string excitations. If the model is valid, this may
hold clues towards understanding back reaction due to Hawking radiation.  A
number of weak spots and open questions regarding the model are also noted.
\Date{July, 1997}
\vfill
\eject

%%%%%%%%%%%%%%%%%%%%%%%%%%%%%%%%%%%%%%%%%%%%%%%%%%%%%%%%%%%%%%%%%%%%%%%%

\newsec{Introduction}

In this paper, we explore further the effective string model 
introduced in \larsen\ for the quantum
microstates of the basic family of 5-dimensional, triply charged black holes 
\refs{\callan,\hms}.  
In one corner of
the black hole parameter space, near extremality and with the momentum 
charge much smaller than the other 
two charges, the quantum mechanical properties of the system, 
entropy and Hawking emission, are well
approximated by a model based on momentum excitations of D-branes 
\refs{\callan,\dmtwo,\ms}. 
We will refer to this as the momentum dominated limit. 
It was proposed in \larsen, however,  that a weakly coupled 
effective string model continues to hold throughout the black hole 
parameter space, for arbitrary combinations
of charges and arbitrarily far from extremality.  

The main evidence presented in \larsen\ for this model comes 
from comparing the thermodynamic properties of the effective 
string with factors appearing in the frequency 
(and charge) 
dependent black hole greybody factors.
The low frequency limit of the black hole greybody factors, or 
equivalently the absorption coefficient, $\sigma_{abs}$,
was calculated in the momentum dominated limit in \ms\ and shown to 
be in striking agreement with the predictions
of the D-brane model\foot{See \refs{\dkt,\kk,\gub,\marika} 
for some qualifications to this.}.  
These calculations were extended to give a
U-duality invariant result for $\sigma_{abs}$ in \refs{\hawking,\dkt}, 
though still in the near-extreme limit with 
at most one charge small.
%\foot{Comments about U-duality related limits here.}.  
It is agreement between these latter results and an appropriate limit of the  effective 
string model in \larsen\ 
that was cited as evidence for the model.  The results of \ms\ were also 
extended to the case of two small charges
in \km, with a further obvious U-duality invariant extension, noted in \dkt, giving 
improved agreement
with the effective string model of \larsen.
We will argue below that this series of 
increasingly precise results for the low frequency limit of $\sigma_{abs}$ points 
towards an ultimate U-duality invariant expression, 
which would hold throughout the black hole parameter space.  This is then an appropriate 
setting to test the effective
string model of \larsen\ and we find exact agreement between the expressions.

A number of further results are presented here. 
 We give a calculation of $\sigma_{abs}$ for scalar fields
carrying arbitrary combinations of the three charges.  The presumed extension of this 
result over the black hole 
parameter space again agrees with the effective string model of \larsen .  
We observe that the spacetime absorption coefficient $\sigma_{abs}$ has the structure of 
an absorption rate which would result from the interaction of two
Bose-Einstein distributions of massless, triply charged
particles which we label `$R$' and `$L$'  which live in a 1+1 dimensional space of radius
$R_{eff}$.   Each distribution is characterised by temperatures
$T_R$ and  $T_L$ and charge potentials $\Phi _i ,\  i=1,2,3$.
Species $R$ and $L$ interact to absorb or emit a third massless boson.
The thermodynamic parameters inferred from $\sigma_{abs}$ - $T_L , T_R$, 
and $\Phi _i $ - agree exactly with the parameters of the
effective string thermodynamic model throughout the parameter space.  Futher,
from the thermodynamics of the effective string model, one can extract the 
effective radii $R_{L,R}$ for the one dimensional spaces on which the left and 
right moving excitations move.
We show that these three lengths are all the same, $R_R=R_L=R_{eff}$.

%We will argue that there is suggestive, and hopefully informative, agreement
%between the spacetime absorption coefficient, and the general thermodynamic 
%model presented in \larsen and further explored here. First, we will present
%a conjecture for the exact absorption coefficient at low energies. The
%expressions for $\sigma_{abs}$ which have been computed all have to assume
%that the spacetime is near extremal. Below we will argue for an
%expression which ``includes all the $r_o$ terms''. Secondly, we repeat
%the spacetime calculation for $\sigma_{abs}$ for a scalar field which carries all three
%charges. 

If the effective string model of \larsen\ is correct, we may be able to learn something
interesting about the back reaction due to Hawking radiation. Recall that in the
momentum dominated limit, the total mass of the
black hole is accounted for by the sum of the masses of the D-branes,
plus the energies of the left and right moving excitations 
$E_L +E_R$. In the general
case discussed here, one also has $M=E_L +E_R +m_{bkgr}$, where
$E_{L,R}=N_{L,R}/\reff$ are the energies of the interacting L and R
modes, and $m_{bkgr}$ is an additonal ``background mass''. In
an emission or absorption process, we find that all of these
energies change, and that 
\eqn\masschange{|\Delta M |   > |\Delta (E_L +E_R )| ,}
over the range of parameter space which we have checked. In the D-brane
picture this would correspond to, e.g., a change in the number of
5-branes and 1-branes, and hence in the effective radius. It would
be of interest to understand what this corresponds to in the
black hole picture.

%This paper is organized as follows. ****

\newsec{Black Holes and Greybody Factors}
Hawking showed that the energy spectrum emitted by a black hole is given by
\eqn\emit{{dE\over dt}(\omega,q_i)={\omega\sigma_{abs}(\omega,q_i)\over 
2\pi (e^{(\omega-q_i\Phi_i)/T_H}-1)},}
where $\omega$ and $q_i$ are the emited energy and charge, $T_H$ and 
$\Phi_i$ are the Hawking temperature
and chemical potential and $\sigma_{abs}$ is the spacetime absorption coefficient for the 
mode, as required by detailed balance.
Since $\sigma_{abs}$ modifies the black body form of the emitted spectrum, 
it is known as a ``greybody'' 
factor.

The $5$-dimensional metric and gauge fields are
\eqn\metric{\eqalign{ds^2&=-f^{-2/3}hdt^2+f^{1/3}\left(h^{-1}dr^2+r^2d\Omega^2\right),\qquad
A_{ti}={\mu\sinhi\coshi\over f_i r^2},\cr
f&=\prod_{i=1}^3 f_i,\qquad f_i=1+{\mu\sinh^2\delta_i\over r^2},\qquad h=1-{\mu\over
r^2}.\cr}} In most of what follows, we will follow \larsen\ and 
set the string coupling and the sizes of the internal 
dimensions\foot{The 10-dimensional and 5-dimensional gravitational couplings are given
by $\kappa_{10}^2=8\pi G_{10}=64\pi^7g^2$ and $\kappa_5^2=8\pi G_5={2\pi^2 g^2\over RV}$.}
to one, 
$g=R=V=1$.
The mass $M$, charges $Q_i$ (i=1,2,3) and entropy $S$ of the black holes 
can then be expressed in terms of three boost parameters $\delta_i$ 
and a nonextremality parameter 
$\mu$ as\foot{This notation for the charges simplifies the formulas.
We will sometimes, however, refer to the charges $Q_i$, $i=1,2,3$ as the $1$-brane, $5$-brane and
momentum charges, respectively.  We will also make use of the notation $\mu=r_0^2$,
$r_i=r_0\sinh\delta_i$.}
\eqn\basics{M = \half\mu\sum_i\cosh 2\delta_i,\qquad Q_i = \half \mu\sinh 2\delta_i,\qquad
S=2\pi\mu^{{3\over 2}}\prod_i\cosh\delta_i .}
The inverse Hawking temperature and chemical potentials are given by
\eqn\thermo{\beta_H = 2\pi\mu^\half\prod_i\coshi ,\qquad \Phi_i=A_{ti}(\mu)=\tanh\delta_i}

First, consider scattering by an uncharged  scalar field satisfying 
$\nabla^2\Phi=0$ and restrict to the S-wave sector
$\Phi=e^{-i\omega t}\phi(r)$.  Then in terms of the radial coordinate 
$v={r_0 ^2 \over r^2}$, 
\eqn\waveinv{(1-v){d\over dv}\left( (1-v) \phi '(v) \right) +
\left[ C_0 +{C_1\over v} +{C_2 \over v^2 } +{C_3 \over v^3 } \right]\phi =0 ,}
where the coefficients $C_k$ are given by
\eqn\constants{\eqalign{
C_3&={\omega^2\mu\over 4},\qquad C_2={\omega^2\mu\over 4}(\sum_i\sinh^2\delta_i),\cr
C_1&={\omega^2\mu\over 4}(\sum_{i<j}\sinh^2\delta_i\sinh^2\delta_j),\qquad
C_0={\omega^2\mu\over 4}\prod_i \sinh^2\delta_i .\cr}}
A scalar carrying momentum charge also satisfies an equation of this form 
\refs{\ms,\igor,\hawking,\dkt}. 
We show below that this form of the equation, with different coefficients $C_k$, 
holds in the case of a scalar field carrying general values of the three charges as well.

As discussed in the introduction, a sequence of continually improving results for 
the absorption coefficient $\sigma_{abs}$ have appeared in the literature
\refs{\dmtwo,\ms,\hawking,\dkt,\km}.  The first of these references 
\dmtwo\ gives the leading term in the power
law expansion of $\sigma_{abs}$ in frequency $\omega$, which is proportional to the
area of the black hole horizon (see Appendix and \dgm).  
The later calculations, starting with \ms,  all give a form for the absorption 
coefficient (still in a low frequency approximation)\foot{We note that
the expressions of this form which have appeared in the literature,
are not strictly consistent in keeping powers of the small
parameter $\epsilon$ ($\epsilon =\omega\ max\{r_i\}$ for neutral emission). One
problem, noted in \km, is that to get
the form of the ratio of Bose factors in (8), gamma
functions have to be expanded as {\it e.g.} $\Gamma ( 1- C_2 )\approx \Gamma (1) $,
so that terms of order $\epsilon ^2$ are dropped. However, the exponentials
in (8) contain all powers of $\epsilon$, and hence have been
selectively kept. Also, in the matching procedures used in \km  , the 
second Bessel function has been simply dropped.  However, this can and 
should be incorporated, and leads to additional order $\epsilon ^2$ corrections
in the prefactor of (8). Again, dropping these terms while keeping
all the terms in the exponential function is inconsistent.  
The justification here for studying the form (8) for $\sigma_{abs}$ is that the Bose
factors have central physical significance. This constitutes an educated guess
about which higher order terms to keep.}.
\eqn\absorb{\sigma_{abs} = \pi^2 \omega^2 \mu a_L a_R
 {e^{2\pi(a_L+a_R)}-1\over (e^{2\pi a_L}-1)(e^{2\pi a_R}-1)},}
with the constants $a_{L,R}$ determined in terms of the coefficients  
$C_k$ in the wave equation \waveinv .

Maldacena and Strominger \ms\ worked directly in the momentum dominated, near extreme limit
\eqn\dlimit{r_0,r_3\ll r_1,r_2,\qquad r_0^3\ll r_1r_2r_3 .} 
and by matching solutions of the hypergeometric equation near the horizon to 
solutions of Bessel's equation
near infinity found 
\eqn\alrd{a_{L,R}=\sqrt{C_0+C_1}\mp\sqrt{C_0},} 
with $C_0,C_1$ approximated in the limit \dlimit.
In \refs{\hawking,\dkt} it was noted that the restriction to the limit 
$r_3\ll r_1,r_2$
was not necessary, and that, so 
long as {\it at most} one of the charges was small, \alrd\ holds with the 
{\it exact values} of 
$C_0,C_1$. 
Since each coefficient $C_k$ is symmetric in the three charges, this
gives a U-duality invariant extension of the results of \ms .
Klebanov and Mathur \km\ found an improved mapping of the near horizon regime to the 
hypergeometric equation giving
\eqn\alrkm{a_{L,R}=\sqrt{C_0+C_1+C_2}\mp\sqrt{C_0},}
with the constants $C_k$ evaluated in the limit of two small charges
\eqn\kmlimit{r_0,r_2,r_3\ll r_1,\qquad r_0^3\ll r_1r_2r_3 .}
An obvious U-duality invariant extension of their result holds as well, with the exact 
coefficients 
$C_0,C_1,C_2$ \dkt .

These evolving results for the coefficients 
$a_{L,R}$ in the papers \refs{\ms,\hawking,\dkt,\km} point towards a possible ultimate form 
\eqn\alr{a_{L,R}=\sqrt{C_0+C_1+C_2+C_3}\mp\sqrt{C_0}.}
In terms of the boost parameters this combination simplifies greatly to give 
\eqn\alrexplicit{a_{L,R}={\omega\mu^\half\over 2}
\left(\prod_i\coshi \mp \prod_i\sinhi\right ).}
This form, we will see, fits in well with the effective string model of \larsen.
One virtue of the expression \alr\ is that it has the correct limit as $\omega$ 
goes to zero.
The leading order term in a power law 
expansion of \absorb\ is
\eqn\leading{\sigma_{abs}\simeq {1\over 2}\pi\mu(a_L+a_R)\omega^2, }
which is proportional to the exact black hole area for
$a_{L,R}$ as in \alr .  
On the other hand, the low frequency limit of \alrkm , 
for example, misses $A_H$ by a term of order ${r_0^6\over r_1^2r_2^2r_3^2}$.
It seems likely that the correct result is of the form \absorb\ 
with coefficients \alr\ 
times a function $f(\omega)$, with $f(0)=1$.  
For the  purposes of discussion, we will assume that the low energy 
absorption coefficient has the 
form \absorb\ with 
coefficients \alr\ in some meaningful 
approximation.

\newsec{General Charges}
We have also calculated the absorption coefficient for a scalar field which 
carries an arbitrary 
combination
of the three charges.  Such a generally charged 
scalar does not arise via dimensional reduction of the $10$-dimensional supergravity 
lagrangian, since it does not correspond to a perturbative degree of freedom.  
Only scalars carrying KK charge will arise in this way.  
Rather it is an effective scalar field describing the propagation of
$5$-D particles which come from combinations of strings and $5$-branes wrapped 
around the internal dimensions 
and boosted along the common string.  
The leading coupling of the scalar to the gauge fields is through
the standard gauge covariant derivative.  
The coupling to the $5$-D moduli scalars, which act as a 
mass term, 
can be inferred via U-duality from the KK charged case where the equation 
of motion is known from
dimensional reduction.

We find that the $5$-D, U-duality invariant wave equation for a scalar carrying 
charges $q_i$ is given by
\eqn\genwave{D_\mu D^\mu\phi-f^{2/3}\left(\sum_i {q_i^2\over f_i^2}\right) \phi=0,\qquad
D_\mu\phi=\left(\nabla_\mu-i\sum_j q_jA_\mu^j\right)\phi .}
The mass term displays the weighted sum of squares of the charges appropriate for a BPS 
scalar field.
Taking the scalar $\phi$ to be a function of $r,t$ only leads to an equation of the 
form \waveinv\ 
with a 
rather complicated set of coefficients $C_k$.  However, combinations which give the 
constants $a_{L,R}$ 
according to
\alr\ simplify considerably to give
\eqn\chargedalr{a_{L,R}= \half\mu^\half\left\{
\left( \omega-\sum_i q_i\tanh\delta_i\right)\prod_i\coshi \mp
\left( \omega-\sum_i q_i\coth\delta_i\right)\prod_i\sinhi\right\} }
and hence, using \thermo, we have
\eqn\alrsum{2\pi(a_L+a_R)=\beta_H (\omega-\sum_iq_i\Phi_i).}
We will see below that this gives an absorption rate in agreement with the 
effective string model.

\newsec{Modeling the Absorption Coefficient}
Before turning to the model of \larsen, we want to see what general features of the
presumed black hole microstates are suggested by the low energy form \absorb\ of the
absorption coefficient $\sigma_{abs}$.  In particular, we find that the form of $\sabs$
is consistent with a picture of left and right moving Bose gases confined to a 
compact one-dimensional
space, which can interact to emit or absorb excitations, which we will call loops, 
propogating in the bulk of spacetime.  In addition, we can infer the 
effective radius $R_{eff}$ of
the one-dimensional space.
%The picture of two noninteracting  ideal gases of massless, triply charged
%bosons, which live on a circle of length $L_{eff}$ and can emit or absorb
%another boson (the loop), comes from
%studying the spacetime absorption coefficient \absorb . 
Consider the expression \absorb\ to have the form $\sigma_{abs}= P\sbabs$, where 
\eqn\factorinto{P=\pi^2\omega^2\mu a_La_R,\qquad 
\sbabs={(e^{2\pi(a_L+a_R)}-1)\over(e^{2\pi a_L}-1)(e^{2\pi a_R}-1)}.}
Then the combination of exponential factors $\sbabs$ has a statistical interpretation in
terms of a pair of
Bose distributions and the prefactor $P$ contains information about the energy 
dependence of the interaction
vertex and the dimension and size of the space on which the excitations move.

Consider uncharged emission.  
Let $\rho_{L,R}(\omega_{L,R})=1/(e^{\beta_{L,R}\omega_{L,R}}-1)$ be the distributions for
two Bose gases at inverse temperatures $\beta_{L,R}$.  
We suppose that a left-mover and a right-mover, with $\omega_L=\omega_R=\omega/2$,
can annihilate to form a loop of energy $\omega$ and that the reverse process 
in which a loop is converted to a left and right moving pair is also
possible.
The net absorption coefficient 
is the difference between the microphysical probabilities for the
system to absorb or emit a loop.  These two processes have different Bose enhancement
factors which combine to give the form of $\sbabs$. 
Fix the initial state
to have $l$ incident loops, then to leading order in
the coupling between the two gases, the relevant statistical factors are
%To leading order in the coupling, the brane can either
%abosorb or emit a loop. These two processes have different Bose enhancement
%factors, which implies that the net absorption rate over a time scale 
%long compared to the time scale for individual interations is  proportional to
\eqn\bose{l \left[\rho_L(\omega/2)+1\right]\left[\rho_R(\omega/2)+1\right]
- (l+1)\rho_L(\omega/2)\rho_R(\omega/2)}
In the case where the number of loops is large $l\gg 1$, as is necessary
for the classical limit, $l\simeq l+1$.  The number of loops $l$ then factors out and
will ultimately be divided out when normalizing by the incident flux, yielding the factor
\eqn\bosetwo{ {e^{\beta_L\omega/2+\beta_R\omega/2}-1\over (e^{\beta_L\omega/2}-1)
(e^{\beta_R\omega/2}-1)},}
which matches\foot{This was noted in \km\ without including the loops.
Putting the loops in a coherent state with high occupation number gives the same result.}
 the form of $\sbabs$, if we identify 
\eqn\identify{2\pi a_{L,R}=\beta_{L,R}\omega/2.}

%So the second factor in the spacetime absorption coefficient \absorb is
%a statistical factor (noted in \km without the loops). Writing out 
%$a_L , a_R$ as in \dkt ,\hawking , one can read off the  three 
%``U-dual temperatures" when momentum excitations, one-brane, or
%five-brane excitations dominate. That is, the form of the spacetime 
%coefficient is very suggestive for the existence of three such 
%corners in the parameter space, even though we only have an
%explicit vertex interaction in the case of the momentum modes.

%There is more information in $\sigma$. The prefactor $\pi ^2 \omega ^2 r_o^2 
%a_L a_R $ contains information about the space on which the interacting 
%degrees of freedom live. Suppose that the spacetime absorption coefficient
%summarizes a microphysical interaction which is of the form
%``$L +R\rightarrow loop$'' and the reverse process. 

In order to understand the physical information in the prefactor $P$, we need to
work with a more detailed model.  Assume that the two interacting species, which we've
labeled `$L$' and `$R$', live on a d-dimensional subspace of the compact
dimensions with volume $\vin $.
%we think of this space as contained in the D-branes.
Assume also that the degrees of freedom have an associated d+1-momentum, which is
conserved in the interactions, \ie\ 
$p_L ^a +p_R ^a =p_{loop} ^a$ in an interaction, where $a$ is a direction tangent to
the space on which the excitations move.
Following the conventions of \dmtwo, write the
interaction vertex in the form 
\eqn\vertex{\kappa_{10}^2 \sqrt{2} (2\pi )^2 (-i A),}
where the amplitude $A$ is left general\foot{In the momentum dominated limit, 
the amplitude is
given by $|A|^2 =(p_L \cdot p_R)^2 = 4\omega _L^2 \omega _R^2$.}. 
If left and right movers  with $\omega_{L}=\omega_R=\omega/2$ and 
$\vec{p}_L=-\vec{p}_R$ combine to give a
neutral loop of energy $\omega$,
then Fermi's Golden Rule
gives the rate for emission of a loop from the system to be
\eqn\emrate{ \Gamma _{em}d\omega = { 1\over 2\pi}\kappa _5 ^2 {\vin\over (2\pi)^{2d} }
 {|A|^2 \over\omega _L \omega_R\omega } \rho _L (\omega _L )
\rho _R (\omega _R ) (\rho _l (\omega ) +1 )d^4 k. }
The net absorption coefficient is again the difference
between the microphysical absorption minus emission rates, normalized 
by the number of incident loops. For $l\gg 1$, 
%as in passing between \bose\ and \bosetwo, 
one then finds 
\eqn\brabs{\eqalign{\sigma _{abs}&= \Gamma _{abs} - \Gamma _{em}\cr 
& = {\kappa _5 ^2 \over (2\pi )^{2d-1}}
{ \vin |A|^2 \over 2}
{e^{\beta_L\omega/2+\beta_R\omega/2}-1\over 
(e^{\beta_L\omega/2}-1)(e^{\beta_R\omega/2}-1)}\cr}}
Now compare the expression for the prefactor $P$ in \brabs\  
to that for the spacetime
absorption coefficient in \factorinto .  
%In order to learn about the energy dependence
%of the amplitude $A$ and the dimension $d$ of the space $\vin$, 
%it is sufficient to consider neutral emission. 
Plugging the explicit
expressions \alrexplicit\ for $a_{L,R}$ into the prefactor $P$ in \factorinto\ gives
\eqn\stpre{\eqalign{ P = & {\pi ^2 \over 4}\omega ^4 \mu^2
\left(\prod_i\cosh^2\delta_i-\prod_i\sinh^2\delta_i\right) \cr
 & \equiv{\kappa_5^2\over 2\pi}(2\pi R_{eff}){\omega^4\over 8}  \cr}}
%
%{\pi ^2 \over 4}\omega ^4 \left( r_1 ^2 r_5 ^2 +r_1 ^2 r_n ^2 +
%r_5 ^2 r_n ^2 +r_o ^2 (r_1 ^2 + r_5 ^2 +r_n ^2 ) +r_o ^4 \right) \cr
%
%= & {\pi ^2 \over 4}\omega{g^2 \over RV} ^4 \reff  \cr}}
%
The last equality above, which has been written in a way to facilitate comparison 
with \brabs, 
serves as a definition of $\reff$.   Recall that the $5$-dimensional 
gravitational coupling $\kappa_5^2$ has the dimension of length cubed, 
so that the units work
out to give a one dimensional volume (\ie\  $d=1$) $V_{in}$ proportional to $2\pi\reff$.   
Similarly, we see that the amplitude $A$ has energy dependence proportional to $\omega^2$.
There is an overall undetermined constant in the determinations of $A$ and $V_{in}$ , which 
may
be fixed by looking in the momentum dominated limit.  If we take
\eqn\determined{V_{in}=2\pi\reff,\qquad |A|^2={\omega^4\over 4} .}
then the effective radius is given by
\eqn\radius{R_{eff}=\mu^2\left(\prod_i\cosh^2\delta_i-\prod_i\sinh^2\delta_i\right),}
and in the momentum dominated limit $\reff$ reduces to 
\eqn\effrad{\reff\simeq RN_1N_5,}
where, in the notation of \hms,  $N_{1,5}$ are the number of one-branes and five-branes.
This is the `fat black hole' result of \refs{\dmone,\msuss}.
We will see below
that the same full value of the effective radius $\reff$  in \radius\ emerges 
in the effective string model of \larsen\ as the ratio of the entropy 
to the temperature, in 
the usual thermodynamic relation for a one dimensional gas.

\newsec{Thermodynamics of the Effective String Model}

The effective string  
proposal made in \larsen\ is initially based on the observation that the entropy $S$
of the black hole system given in \basics\  
can be exactly
broken up into the sum of two 
terms\foot{This split is particularly compelling in 
the rotating case, which we do not consider here.}, 
which may be interpreted as the contributions
of right and left handed massless excitations moving in 1-dimension,
\eqn\entropy{
S=2\pi\left(\sqrt{N_R}+\sqrt{N_L}\right),\qquad
N_{L,R}={\mu^3\over 4}\left(\prod_i\coshi \pm \prod_i\sinhi\right)^2 .}
$N_{L,R}$ are interpreted as the excitation levels of the right and left handed sectors of 
a $c=6$ conformal field theory\foot{This comes from 
$S=2\pi\sqrt{(N_B+N_F/2)N_{L,R}/6}=2\pi\sqrt{cN_{L,R}/6}$, where $N_B$ and $N_F$ are
the number of $1$-dimensional bosons and fermions and $c=N_B+N_F/2$.}.
For reference, in the limit where momentum excitations dominate, 
$N_{L,R}=N_1N_5N_{L,R}^{mom}$.
It could be that this split is an arbitrary one and has no physical relevance.
However, evidence that there is something interesting going on comes from a comparison made 
in \larsen\ between the thermodynamic properties of the effective left and right 
movers inferred from $S_{L,R}$ and
the form of the spacetime absorption coefficient.  We will see that this agreement, 
made at an approximate level in \larsen\ based on the results for the
spacetime absorption coefficient presented in \refs{\ms,\hawking,\dkt}, becomes
exact for the form of the spacetime absorption coefficient assumed in section (2).

The thermodynamic properties of effective left and right 
movers derived in \larsen\ are as follows.
The left and right contributions to the entropy are given by
\eqn\lrentropy{S_{L,R}=\pi\mu^{{3\over 2}}\left(\prod_i\coshi \pm \prod_i\sinhi\right )
=2\pi\sqrt{N_{L,R}}.}
Inverse temperatures $\beta_{L,R}$ are computed for the left and right movers as
\eqn\temp{\beta_{L,R}=\left( {\partial S_{L,R}\over \partial M}\right )_{Q_i}
=2\pi\mu^{{1\over 2}}\left(\prod_i\coshi \mp \prod_i\sinhi\right ),}
and chemical potentials are given by
\eqn\chemical{\left(\beta\Phi_i\right)_{L,R}=
-\left({\partial S_{L,R}\over\partial Q_i}\right)_{Q_{j\ne i},M}
=2\pi\mu^\half\left(\tanh\delta_i\prod_{j=1}^3\cosh\delta_j 
\mp \coth\delta_i\prod_{j=1}^3\sinh\delta_j\right ).}

The comparison with the form of the absorption coefficient given in section (2) 
is now straightforward.  In the neutral case we have using \alrexplicit 
\eqn\another{2\pi a_{L,R}= \beta_{L,R}\omega/2,}
and hence the exponential terms in the 
absorption cross section for the effective string calculated as in section 
(4) agree with the
exponential terms in the  
spacetime absorption coefficient.  The generally charged case works similarly.  We have 
from \chargedalr ,
\eqn\yetanother{2\pi a_{L,R}=\half\left[\beta\left(\omega-q_i\Phi_i\right)\right]_{L,R}, }
and again the the exponential terms in the absorption coefficients agree.

In order for the prefactors to agree as well, the radius 
of the effective string must match
with the effective radius deduced from the spacetime absorption coefficient in section (4).
Here, we determine $R_{eff}$ by the observation that the entropy and
energy are both extensive quantities, and so proportional to the length of the string. 
For a one
dimensional ideal Bose gas,
\eqn\extensive{\eqalign{
E_{L,R} =& {c\over 6} \pi ^2 R_{L,R} T_{L,R}^2 \cr
S_{L,R}=& {c\over 6} \pi ^2  R_{L,R}T_{L,R} \cr }}
Given the expressions for $S_{L,R}$ and $\beta_{L,R}$ in equations \lrentropy\ and \temp\ 
one finds
the same effective length for both the $R$ and $L$ gases, $R_L =R_R$ as would
be needed for the consistency of the model, and the value matches that determined from the
prefactor $P$ in \radius ,  
\eqn\effectiver{R_{eff}= \mu^2\left(\prod_i\cosh^2\delta_i -\prod_i\sinh^2\delta_i\right). }
We also note that, in the 
momentum dominated limit 
\eqn\momlim{\delta_3\ll \delta_{1,2} \ \ \  \delta_1,\delta_2\gg 1 }
 we have $\coshi\approx\sinhi$ for $i=1,2$, 
leading to
$R_{eff}= N_1N_5R$ as previously in \effrad .

%A further check on the applicability of this generalized string model
%is to see if the greybody factors match. In the spacetime absorption
%coefficient \absorb  the factors $2\pi a_{L,R}$ appear in the exponents,
%with $a_{L,R}$ given in \chargedalr  . In the string model, these
%factors arise as Bose statistical factors, as explained in Section 4.
%In an interaction $N_{L,R}$ change by $\Delta N_{L,R} $,
%and agreement requires that

Another way to express the Bose statistical factors for the effective string excitations is
in terms of the changes $\Delta N_{L,R}$ in an absorption or emission process.  It follows
from the thermodynamics of the system, that we must have
\eqn\agreeone{2\pi a_{L,R} =\beta _{L,R} {\Delta N_{L,R}
\over R_{eff} }, }
under arbitrary variations of $\mu $
and $\delta _i$. This can be verified by direct calculation.
The sum of these two equations gives a relation which
will be of later use. Using \alrsum\ we have
\eqn\agreetwo{\beta _H (\omega -\Sigma q_i \Phi _i )
={1\over R_{eff}}(\beta _R \Delta N_R +\beta _L \Delta N_L ) }
From the first law,
the left hand side is the change in the entropy of the black hole
when it emits or absorbs energy $\omega$ and charges $q_i$. The
right hand side is the change in the entropy of the brane system,
$\Delta S_R +\Delta S_L$. 
%The equality can be shown by direct calculation,
%but also follows from the construction of the generalized string model,
%in which $S_{bh}= S_L +S_R$ over the entire parameter space.

\newsec{Some Weak Spots or Clues?}
In the last section, 
we found impressive agreement between the thermodynamic properties of the
the effective string model and features of the spacetime absorption coefficients.
In this final section, we would like to point out some other features of the 
proposed correspondence
which do not show such obvious agreement and hence stand as open questions.
We begin by discussing the relationship between the combined
energies of the right and left moving excitations compared to the black hole ADM mass.  

The relations \temp , \extensive\ and \effectiver\ are standard thermodynamic relations
for a pair of $1$-dimensional ideal gases. For some of the thermodynamic quantities
there is a direct relation to analogous properties of the corresponding black hole.
For example, by construction, the entropy of the black hole is 
$S_{bh}= S_R +S _L$, and also the inverse Hawking temperature is
$\beta _H = (\beta _L +\beta _R)/2$.  
We can ask what are the meanings at the spacetime level of the total energies
$E_{L,R}$ carried by the left and right movers?  Combining the formulas above, one finds
the additional standard relation
\eqn\energylr{ E_{L,R}={N_{L,R} \over \reff}. }
However, it is easily checked that 
the sum $E=E_R+E_L$ does not equal the ADM mass of the black hole given in
 \basics  .  In the momentum dominated limit, the difference
between the ADM mass and the sum of the energies carried by the left and right
moving excitations is attributed to the mass of a background soliton on which
they propagate.  In this limit, the excitation energies reduce to $E_{L,R}\simeq
N_{L,R}^{mom}/R$. The mass of the background is then
\eqn\stringmass{\eqalign{m_{bkgr}&=
M-(E_R+E_L)\cr &\simeq {R\over g} N_1 + {RV\over g}N_5,\cr}}
where we have again restored $R,V,g$.  The soliton mass $m_{bkgr}$ is
then the sum of the masses of the $1$-branes and $5$-branes.
Similarly, in the model considered in \km, the energy of the excitations
correspond to the momentum and $1$-brane energies, and the the difference between
these and the ADM mass is just the mass of the
$5$-branes.  In general, however, $m_{bkgr}$ defined according to the first
line of  \stringmass\ does not appear to have a simple interpretation.  
All three types of constituents, momentum, $1$-branes and $5$-branes are 
intertwined in the both the excitations and the background on which they propagate.

This leads to a further interpretational question.
In the momentum dominated limit \momlim , if
$\mu$ is essentially fixed, then when a loop is emitted from the brane,
it is only the excitation energies $E_{L,R}$ which change.  
The mass of the branes
 $m_{bkgr}$ remains unchanged, as one would expect. 
In the general case, however, both the excitation energies and $m_{bkgr}$
change with emission. We have
\eqn\lrchange{\Delta (E_L +E_R ) = {\Delta ( N_L +N_R )\over \reff} +
 (N_L +N_R )\Delta \left( {1\over \reff}\right).} 
Consider neutral emission, in which a left and right mover,
each with energy $\omega/2$ combine to form a loop of energy $\omega$,
and $\Delta N_L=
\Delta N_R=\Delta N$. It then follows from the thermodynamic relations above,
or directly from \agreetwo  , that for neutral emission
\eqn\masschange{\delta M= \omega= {2 \Delta N\over R_{eff}} .}
and therefore from \masschange\ and \lrchange
\eqn\masschtwo{\Delta M =\Delta(E_L +E_R )-(N_L +N_R )\Delta 
\left( {1\over \reff } \right) }
So if we let $M=E_L +E_R +m_{bkgr}$, when the ADM mass changes, in addition
to a change in $E_L +E_R$, there is a change in the background mass of
$\Delta m_{bkgr}= (E_L +E_R ){\Delta \reff \over \reff}$.

It is of interest to know what the sign of this extra change is. While
it is straightforward to write down the variation of $\reff$,
it seems difficult in general to determine it's sign because
$\Delta \delta _i$ and $\Delta \mu$ can have general signs. However
for neutral emission the condition that the charges $Q_i$ are
fixed relates the variations. In this case, one can check the
behavior of the variation in $\reff$ in various limits. Checking for
equal $\delta_i$, one large $\delta _i$, and two large $\delta _i$,
we find
\eqn\chaninr{\Delta {1\over \reff} =-2 {\Delta\mu\over \mu}
{1\over \reff} \alpha \  ,\  0\leq \alpha\leq 1 }
where $\alpha$ is a function of the $\delta _i$ and is found to be a positive
number, between zero and one, in the above mentioned limits.
Since for neutral emission, $\Delta \mu$ has the same sign
as $\Delta M$, this means that
\eqn\bigchange{|\Delta M | >|\Delta (E_L +E_R ) | }
Of the cases checked the largest value for $\alpha$, and hence the
largest additional contribution to the change in the ADM mass,
was for equal $\delta _i$ or one large $\delta _i$, which gives
$\alpha$ close to one.

In the effective string picture, we interpret equation \bigchange\   as
saying that in an emission process, there is some adjustment of
the background soliton structure, which leads to an additional
contribution to the total emitted energy (and likewise for
absorption). An interesting question is what
does this mean in the black hole spacetime? That is, if the
generalized string model is correct, can we learn something
about the backreaction in Hawking emission? Is there also a  division
of field energies in the spacetime, corresponding to the 
division between $E_L +E_R$ and $m_{bkgr}$? Does the additional
emitted energy come from inside or outside of the horizon?

Another area which needs to be explored more fully is how charge is carried by the
effective string excitations.  For agreement with the emission of generally charged scalars,
the effective string excitations must have statistical distributions of the form
\eqn\chargeddist{\rho_{L,R}(\omega,q_i)=
{1\over e^{\beta_{L,R}\omega-(\beta\Phi_i)_{L,R}q_i}-1}.}
In particular, it is clear that the excitations 
must carry all three varieties of charge.  This raises a number of
questions.  Is momentum of the left and right moving excitations along the string still to
be identified with Kaluza-Klein charge, as in the momentum dominated limit?  If so, then 
the U-duality invariance of the model is compromised.  If not, to 
what does this momentum now correspond?  Without an understanding of how charge is
carried by the excitations, the effective string model of \larsen\ 
is incomplete.   Rather, it seems to be a set of interesting
thermodynamic  relations still in search of a microscopic model.

Finally, it remains to be seen whether the effective string model of \larsen\ can 
be made to agree with results for the emission of fixed scalars \refs{\kol,\fixed,\fixedtwo,
\kk,\marika}, intermediate scalars \krt , higher angular momentum \refs{\mstwo,\gub} and 
higher spin modes \gubthree .

\bigskip
\noindent
{\bf Note Added:} After this work was complete, the papers \refs{\clone,\cltwo} appeared
which have some overlap with our results.  
In particular, using a different radial coordinate
in the near horizon region, these authors were able to show \clone\ 
that the absorption coefficient
has the form \alr\ which we have conjectured.  They also extract the effective string radius
from the prefactor to the absorption coefficient.
\bigskip
\noindent
{\bf Acknowledgements:} We thank Fay Dowker and Roberto
Emparan for helpful discussions
and the  Aspen Center for Physics for its hospitality while this work was being
completed.   This work was supported in part by NSF grant NSF-THY-8714-684-A01.

\appendix{A}{$\sigma _{abs}\rightarrow A_H$ at low freqencies}

In this appendix we outline the calculation showing that the low
frequency limit of the absorption coefficient for
a charged scalar field in the spacetime \metric\ is the horizon area,
and derive how low the frequency must be. An analogous result has, of course,
appeared previously for scalar fields in four dimensions. In \dgm\  part
of the argument was given for a black hole in any dimension. However,
conflicting approximations are made at different points in the 
calculation in \dgm, and one must additionally show that there is a parameter range 
over which the result actually holds.  The main result of this appendix are
the conditons (A.11) and (A.12).

We consider the wave equation for a scalar field carrying a single charge in the
background \metric\ written in terms of the tortoise coordinate
\eqn\wavetort{\chi ''(r_* ) +\left[ \omega _{\infty} ^2 -V_{coul} -V_{grav}
\right] \chi =0, }
where  $ \lambda =r^{3/2} f^{1/4}$, $\chi = \lambda \phi$,
$d r_* =\sqrt{f} (1-{r_0 ^2 \over r^2} )^{-1} dr $ , and
\eqn\scatpot{V_{coul} =r_n ^2 {\omega _{\infty} ^2 -\mu ^2 \over r^2 +r_n ^2
},\qquad   V_{grav} ={\lambda '' (r_* )\over \lambda }, }
Here $\omega _{\infty} ^2 =\omega ^2 -k_5 ^2 $ and $\mu =\omega -k_5 (1+
{r_o^2\over r_3^2} )^\half$.

Let $\kappa$ be the surface gravity, $r_1$ be the largest charge in the metric, 
and $k_5$ be the Kaluza-Klein charge. Then near the horizon $r_* \rightarrow
-\infty $, the scattering potentials go to zero exponentially fast, and
\eqn\phihor{\phi \rightarrow A e^{-i\omega _H r_* } ,}
where $\omega _H =\omega -k_5 (1+{r_o ^2 \over r_n ^2 })^{-1/2} $ 
is the frequency of the wave near the horizon. This assumes that
\eqn\approxone{\kappa ^3 r_1 e^{2\kappa r_* }\ll \omega _H ^2 \  ,}
\eqn\approxtwo{ and \ \   e ^{2\kappa r_*} \ll 1 .}
In the asymptotically flat region $r_* \gg r_1$ the solutions are Bessel
functions,
\eqn\phinfinity{\phinf (r ) = \sqrt{{\omega _{\infty} \pi \over 2}}
{e^{-i\pi/4}\over r} \left( H_\nu ^{(2)} (\omega _{\infty} r) +i S H_\nu
^{(1)}(\omega _{\infty} r) \right),\quad
\nu=1-\omega_\infty^2(r_1^2+r_5^2+r_0^2)+\mu^2r_n^2. } 

Since there is no overlap between these two regions, we use a third
region to ``bridge the gap''. The wave equation can be solved
exactly when $\omega _{\infty}=\mu =0$. One can then do a double
power series expansion for $\phi$ in these two parameters. The 
expansion is valid in a ``middle region'', away from $r_* \rightarrow
\pm \infty$. One finds
\eqn\phimid {\phi _o =B[ ln(r- r_o ) +ln (r+r_o ) -2lnr ]
+C . }
To match the near horizon solution \phihor\  to the middle solution
one expands the plane wave, which requires
\eqn\approxthree{\omega _H |r_* |\ll 1 }
in the matching region. Note that the condition \approxone\ requires
a {\it large} negative value of $r_*$, while \approxtwo\ requires
a {\it small} value of $|r_*|$.
The large argument expansion of the near horizon
solution can be matched onto the middle solution, which then can be matched
onto the small argument expansion
of the Bessel function can be matched onto the middle solution;
see \dkt\  for details. The latter expansion requires
\eqn\approxfour{\omega _{\infty} r \ll 1 .}
Matching gives the the coefficients $A$ and $S$. The absorption coefficient
can be computed either as $1-|S|^2$, or as the ratio of the flux of the 
field $\phi$ crossing the horizon to the incident flux from infinity.
Either way gives
\eqn\abscoeff{ \sigma _{abs}= {\omega _H \omega _{\infty}^2 \over
4\pi} A_H .}

Finally, let us combine the conditions for validity of the result
\abscoeff  . Choose values for $\kappa ,r_1$. Then \approxthree\ and
\approxtwo\  require that 
\eqn\condone{\omega _H \ll 2\kappa .}
Now, if 
$-\infty < ln(\kappa r_1 ) <1 $, one can check that \approxone\  is already
satisfied. So it is sufficient to satisfy \condone  .
If $ln(\kappa r_1 )>1$, in order to satisfy \approxone  , one needs
\eqn\condtwo{\omega _H \ll {2\kappa \over ln(\kappa r_1 )},}
which is more stringent than \condone  .

\listrefs
\end